\documentclass[prb,superscriptaddress,twocolumn,showpacs,preprintnumbers,amsmath,floatfix]{revtex4}
\usepackage{amssymb}

\usepackage{graphicx}

\begin{document}

\title{Itinerant Nature of Magnetism in Iron Pnictides: A first principles study}

\author{Yu-Zhong Zhang}
\affiliation{Institut f\"ur Theoretische Physik,
Goethe-Universit\"at Frankfurt, Max-von-Laue-Stra{\ss}e 1, 60438
Frankfurt am Main, Germany}
\author{Ingo Opahle}
\affiliation{Institut f\"ur Theoretische Physik,
Goethe-Universit\"at Frankfurt, Max-von-Laue-Stra{\ss}e 1, 60438
Frankfurt am Main, Germany}
\author{Harald O. Jeschke}
\affiliation{Institut f\"ur Theoretische Physik,
Goethe-Universit\"at Frankfurt, Max-von-Laue-Stra{\ss}e 1, 60438
Frankfurt am Main, Germany}
\author{Roser Valent{\'\i}}
\affiliation{Institut f\"ur Theoretische Physik,
Goethe-Universit\"at Frankfurt, Max-von-Laue-Stra{\ss}e 1, 60438
Frankfurt am Main, Germany}

\date{\today}

\begin{abstract}

Within the framework of density functional theory we investigate the
nature of magnetism in various families of Fe-based superconductors.
(i) We show that magnetization of stripe-type antiferromagnetic
order always becomes stronger when As is substituted by Sb in
LaOFeAs, BaFe$_2$As$_2$ and LiFeAs. By calculating Pauli
susceptibilities, we attribute the magnetization increase obtained
after replacing As by Sb to the enhancement of an instability at
$(\pi,\pi)$.  This points to a strong connection between Fermi
surface nesting and magnetism, which supports the theory of the
itinerant nature of magnetism in various families of Fe-based
superconductors. (ii) We find that within the family LaOFe$Pn$
($Pn$=P, As, Sb, Bi) the absence of an antiferromagnetic phase in
LaOFeP and its presence in LaOFeAs can be attributed to the
competition of instabilities in the Pauli susceptibility at
$(\pi,\pi)$ and $(0,0)$, which further strengthens the close
relation between Fermi surface nesting and experimentally observed
magnetization. (iii) Finally, based on our relaxed structures and
Pauli susceptibility results, we predict that LaOFeSb upon doping or
application of pressure should be a candidate for a superconductor
with the highest transition temperature among the hypothetical
compounds LaOFeSb, LaOFeBi, ScOFeP and ScOFeAs while the parent
compounds LaOFeSb and LaOFeBi should show at ambient pressure a
stripe-type antiferromagnetic metallic state.

\end{abstract}

\pacs{74.70.-b 74.25.Ha 74.25.Jb,71.15.Mb,71.15.Pd}

\maketitle

\section{ Introduction}
\label{sec:one}
After the discovery of the first high-$T_c$ iron-based
superconductor La[O$_{1-x}$F$_x$]FeAs~\cite{Kamihara08} (denoted as
1111 compound), the superconducting transition temperature was
rapidly raised up to 55~K with substitution of La by
Sm~\cite{Ren08}. While various other families of Fe-based
superconductors were reported afterwards, like the 122 compounds
$AE$Fe$_2$As$_2$
($AE$=Ca,Sr,Ba)~\cite{Torikachvili08,Sasmal08,Chen08,Rotter08PRL,Alireza09,Kimber09,Sefat08},
the 111 compounds $A$FeAs
($A$=Li,Na)~\cite{Wang08,Tapp08,Pitcher08,SJZhang09,Parker09,Chen09}
and the 11 compounds Fe$Ch$
($Ch$=Se,Te)~\cite{Hsu08,Mizuguchi08,Yeh08}, the transition
temperature has always been lower than the highest one observed in
the 1111 systems. Besides the continuous experimental attempts to
pursue higher superconducting transition temperatures in the
Fe-based compounds and deeper understanding of high-$T_c$
superconductivity~\cite{Yuan09}, a great effort to understand the
origin of their phase diagram has also been made on the theoretical
side.

Although it is widely believed that magnetically mediated rather
than phonon-mediated pairing dominates the superconducting state due
to its proximity to a stripe-type antiferromagnetic phase~\cite{
Singh08,Haule08,Boeri08,Mazin08,Mazin09,Wang09,Chubukov08,Stanev08,Kuroki08,Graser08,Graser09,Qi08,Yao09,Sknepnek09,Si08,Yildirim08,Seo08,Ishida09},
the origin of the magnetism is still highly under debate. Some of
the experimental work and theoretical studies based on density
functional theory (DFT) support an itinerant scenario of magnetism
due to the fact that the electron and hole sheets of the Fermi
surface are nearly
nested~\cite{Mazin08,Hsieh08,Fink09,Cvetkovic09,Dong08,Opahle08,YZZhang09,Yaresko09}
and correlation effects are not very strong, resulting in a metallic
state of the parent
compounds~\cite{Kroll08,Skornyakov09,Yang09,Aichhorn09}. In
contrast, some authors favor a localized
picture~\cite{Haule08,Si08,Yildirim08,Han09A,Craco08,Ma08,Ma09,Krueger09}
since DFT calculations fail to reproduce the experimentally observed
band splitting in the stripe-type antiferromagnetic
phase~\cite{LYang09}. In these studies the observed small magnetic
moment is attributed to highly frustrated superexchange interactions
which explain the observed low energy spin excitations
well~\cite{Schmidt09}. Apart from the opposing viewpoints above,
various other interpretations co-exist, such as those that propose
that the magnetism could come from the local Hund's rule
coupling~\cite{Johannes09} or from the coexistence of localized and
itinerant electrons~\cite{JWu08,JWu09,Kou08,FWang09}. A recent LDA+U
calculation explains the small magnetic moment by formation of
magnetic multipoles~\cite{Cricchio09} while LDA+DMFT (dynamical mean
field theory) calculations in conjunction with angle resolved
photoemission (ARPES) experiments suggest that involvement of
non-local fluctuations may be crucial~\cite{Barriga09}. Further, a
recent DMFT calculation stressed the importance of the interplay
between frustrated and unfrustrated bands within a two-band Hubbard
model at half-filling~\cite{Lee09}.  Among these theories, it is
presently hard to decide which is the most promising for the
observed magnetism in the parent compounds of the iron-based
superconductors.

In fact, one of the most popular theories mentioned above, the
itinerant scenario of magnetism, was recently substantially
challenged both by experiment~\cite{Xia09} and  density functional
theory calculations~\cite{Moon09}. On the one hand, an ARPES study
on the 11 compound Fe$_{1+x}$Te  shows no evidence of a Fermi
surface nesting at ($\pi,0$)~\cite{Xia09} while magnetic order with
such a wave vector is detected by neutron scattering~\cite{Li09}. On
the other hand, a DFT calculation on LaOFeAs, BaFe$_2$As$_2$ and
LiFeAs based on a pseudopotential method~\cite{Moon09} reveals that
the magnetic moments are enhanced in all compounds by replacing As
with Sb while the Fermi surface nesting with a nesting vector of
($\pi,\pi$) is found with this substitution to be enhanced only in
the 1111 compound but suppressed in the 122 and 111 compounds. These
studies question altogether the applicability of the theory of
itinerant magnetism to the 11, 122 and 111 systems.

The former discrepancy in the 11 compounds was soon resolved by a
new DFT calculation based on the full potential linear muffin tin
orbital (FPLMTO) method which reconciles the theory of itinerant
magnetism with the existing experiments on
Fe$_{1+x}$Te~\cite{Han09B}. It shows that, while the Fermi surface
is nested at ($\pi,\pi$) in the undoped FeTe as in other iron-based
superconductors, doping with 0.5 electrons due to the excess of Fe
in Fe$_{1+x}$Te leads to a strong ($\pi,0$) nesting of the Fermi
surface which corresponds to the observed magnetic ordering.
However, up to now, the second question of enhanced Fermi surface
nesting in 1111 versus suppression in 122 and 111 compounds when As
is substituted by Sb still remains.

In this work, by applying Car-Parrinello molecular dynamics
~\cite{CarParrinello} based on a projector augmented wave (PAW)
basis~\cite{Bloechl}, we will show that, in contrast to the results
from a pseudopotential method~\cite{Moon09}, the magnetic moment and
the Pauli susceptibility at ($\pi,\pi$) are simultaneously enhanced
when As is replaced by Sb in LaOFeAs, BaFe$_2$As$_2$ and LiFeAs,
which strongly suggests that Fermi surface nesting is closely
related to the magnetic moment strength, and the itinerant scenario
of nesting-driven magnetism is still valid in the 111, 122 and 1111
compounds. By further comparing the Pauli susceptibilities of
LaOFe$Pn$ with $Pn$=P, As, Sb, and Bi, we argue that the absence and
the presence of magnetism at ambient pressure in LaOFeP and LaOFeAs,
respectively, originate from the competition between the
instabilities of the susceptibility at ($0,0$) and ($\pi,\pi$),
which again indicates the importance of Fermi surface nesting for
the description of magnetism. We predict that a stripe-type
antiferromagnetic metallic state should be present in the
hypothetical compounds LaOFeSb and LaOFeBi. Finally, we study the
structural and magnetic properties of 1111 compounds including
$RE$OFeAs ($RE$=Ce, Nd, Sm), LaOFe$Pn$ ($Pn$=P, As, Sb, Bi) and
ScOFe$Pn$ ($Pn$=P, As) and predict that LaOFeSb could be a
superconductor with the highest transition temperature among these
compounds.

\section{ Method}
\label{sec:two}

Throughout this paper, the Car-Parrinello~\cite{CarParrinello}
projector-augmented wave~\cite{Bloechl} method is employed to
optimize the lattice parameters and internal atomic positions. These
optimized structures are then used for all subsequent electronic
structure calculations unless stated otherwise. $4\times 4 \times 4$
${\bf k}$-points and doubled ($\sqrt{2}\times\sqrt{2}\times 1$) unit
cells with stripe-type antiferromagnetic order are used when
relaxation of all lattice and electronic degrees of freedom is
performed. We use time steps of 0.12~fs and friction to cool the
systems to zero temperature. Note that the structure optimization is
performed in the magnetic phase. As we shall show below, whenever
experimental structures are available, our optimized structures
compare well with the experimental ones. This is not the case if
structure optimizations are performed within non-spin polarized
calculations as has been frequently pointed out in the
literature.~\cite{Mazin08PRB,Fink09,Opahle08}

We used high energy cutoffs of 408~eV and 1632~eV for the wave
functions and charge density expansion, respectively. The total
energy was converged to less than 0.01~meV/atom and the cell
parameters to less than 0.0005~\AA. Part of our results are
double-checked by the full potential linearized augmented plane wave
(FPLAPW) method as implemented in the WIEN2k code~\cite{Blaha} and
full potential local orbital (FPLO) method~\cite{FPLO}. Results are
consistent among these methods. Throughout the paper, the
Perdew-Burke-Ernzerhof generalized gradient approximation (GGA) to
DFT has been used if not specified otherwise, and comparisons with
the results from local density approximation (LDA) were also
performed. In order to determine if the Fermi surface nesting is the
driving force for the low-temperature stripe-type antiferromagnetic
ordering, we calculate the ${\bf q}$-dependent Pauli susceptibility
at $\omega$=0 and Fermi surface cuts at different $k_z$ planes
without magnetization. These calculations were performed with the WIEN2k
code using $RK_\text{max}=7$. While 40000~${\bf k}$ points in each $k_z$
plane are used in calculating Fermi surface cuts, a three
dimensional grid of $128\times128\times128$ ${\bf k}$ and ${\bf q}$
points and the constant matrix element approximation are employed
for the susceptibility.  For the calculations including Ce, Nd, and
Sm atoms in the nonmagnetic phases, we apply the open core
approximation for the localized $f$ electrons. All calculations were
performed in the scalar relativistic approximation, which usually
provides a good description of structural properties even for
heavier elements~\cite{Eschrig04}. Thus, spin-orbit coupling, which
could be potentially relevant especially for Bi
compounds~\cite{Kozlova05,Wosnitza06}, is neglected in the
calculations for the valence electrons. However, since most of the
weight of the Bi 6p states is well below the Fermi energy and
irrelevant to magnetic ordering, we would expect only minor
modifications for the resulting Fermi surfaces and susceptibilities.

\begin{figure}[tb]
\includegraphics[width=0.35\textwidth]{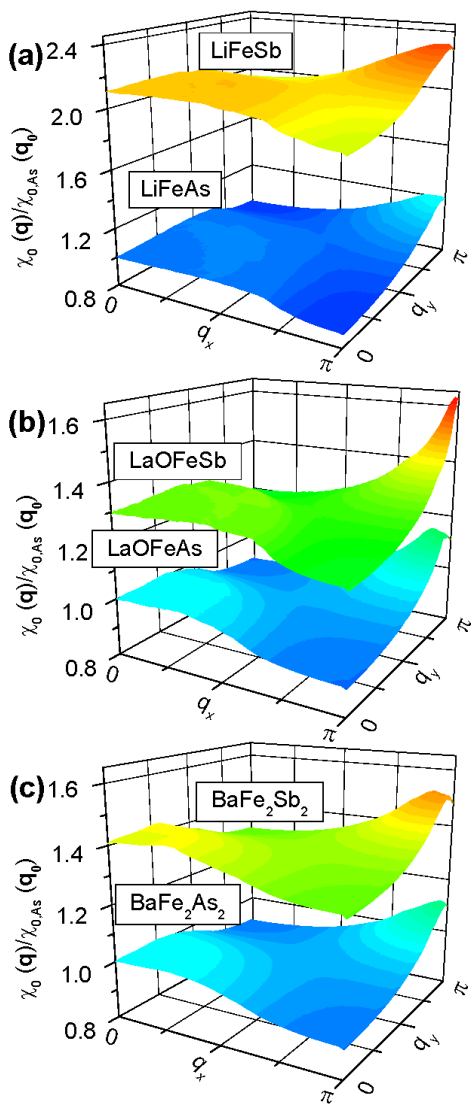}
\caption{(Color online) Comparison of normalized static ${\bf
q}$-dependent
  Pauli susceptibilities at fixed $q_z=\pi$ between arsenide and
  antimonide of (a) 111 compounds, (b) 1111 compounds and (c) 122
  compounds. The normalization factors are the susceptibilities of
  the corresponding arsenide systems for each type of compound at ${\bf
    q}_0=(0,0,\pi)$. Please note that the peak position is not exactly
  at ${\bf q}_{\pi}=(\pi,\pi,\pi)$ since the electron and hole Fermi
  surfaces are nearly nested rather than
  perfectly nested. Here, GGA is used for the DFT calculations.} \label{fig:SUSAsVsSb}
\end{figure}

\section{ Pauli susceptibilities and magnetism in 1111, 122 and 111 compounds}
\label{sec:three}

As is pointed out in Section \ref{sec:one}, a DFT calculation based
on a pseudopotential method within the SIESTA code~\cite{Moon09}
reveals a disconnection between magnetism and Fermi surface nesting
in 122 and 111 compounds, ({\it i.e.}, while magnetization is enhanced,
the Pauli susceptibility at ${\bf q}=(\pi,\pi)$ which is responsible
for stripe-type antiferromagnetic ordering is suppressed when As is
replaced by Sb in BaFe$_2$As$_2$ and LiFeAs), and therefore
questions the scenario of an itinerant nature of magnetism. From our
spin-polarized GGA calculations for LaOFe$Pn$, BaFe$_2Pn_2$ and
LiFe$Pn$ ($Pn$ = As, Sb), the same trends in magnetism are detected as
observed in Ref.~\onlinecite{Moon09}: the ground states are all
found to be stripe-type antiferromagnetic metallic states and
magnetic moments increase with the substitution of As by Sb in
LaOFeAs, BaFe$_2$As$_2$ and LiFeAs.

\begin{table}[!h]
\caption{Comparison between different DFT codes of the structures of
LiFeAs optimized within GGA. $z_{Li}$=0.3385 and $z_{As}$=0.7688 are
obtained from our CP-PAW calculations.}
\begin{tabular}{cccccccccc}
\hline\hline
& $a(\text{\AA})$ & $b(\text{\AA})$ & $c(\text{\AA})$ & $m(\mu_B)$ & $d_\text{Fe-As}(\text{\AA})$ \\
\hline
SIESTA~\cite{Moon09} & 5.482 & 5.285 & 6.190 & 2.54 & 2.434 \\
VASP~\cite{ZLi09} & 5.408 & 5.294 & 6.237 & 1.5 & 2.359 \\
WIEN2k~\cite{YFLi09} & - & - & - & 1.58 & 2.382 \\
CP-PAW & 5.422 & 5.307 & 6.255 & 1.56 & 2.385 \\
\hline
\end{tabular}\label{T1LiFeAs}
\end{table}

However, the magnetic moments we obtained are 1.6~(2.3)~$\mu_B$ in
LiFeAs~(LiFeSb), 2.0~(2.5)~$\mu_B$ in
BaFe$_2$As$_2$~(BaFe$_2$Sb$_2$) and 1.8~(2.2)~$\mu_B$ in
LaOFeAs~(LaOFeSb)~\cite{GGA}, which are notably smaller than those
obtained from the pseudopotential method~\cite{Moon09,Moon08}.
Further comparing the optimized lattice structures, we find that,
while our results are in good agreement with previous GGA
calculations, such as LiFeAs calculated with VASP~\cite{ZLi09} and
WIEN2k~\cite{YFLi09}, there are large differences between our
results and those of Refs.~\onlinecite{Moon09,Moon08} as shown in
Table~\ref{T1LiFeAs}. Furthermore, in Table~\ref{T2BaFeAs}, we show
the comparison between experimental and optimized structural data
for BaFe$_2$As$_2$, where we find that our optimized structure
agrees with the experimental one better than that from
Ref.~\onlinecite{Moon09}. Since the electronic band structure close
to the Fermi level is sensitive to the lattice
structure~\cite{Singh08,Vildosola08}, the conclusion of
Ref.~\onlinecite{Moon09} based on their optimized structures that
there is no connection between Fermi surface nesting and magnetism
is questionable. Therefore, we reinvestigate the nesting property of
the Fermi surface.

\begin{table}[!h]
\caption{Comparison between the experimental structure of
  BaFe$_2$As$_2$ and the optimized structures from different DFT codes
  within GGA. The magnetic moment on each Fe is also shown.
  $z_{As}$=0.6495 is obtained from our CP-PAW calculations.}
\begin{tabular}{cccccccccc}
\hline\hline
& $a(\text{\AA})$ & $b(\text{\AA})$ & $c(\text{\AA})$ & $m(\mu_B)$ & $d_\text{Fe-As}(\text{\AA})$ \\
\hline
Exp.~\cite{Huang08} & 5.616 & 5.571 & 12.943 & 0.87 & 2.392 \\
SIESTA~\cite{Moon09} & 5.756 & 5.590 & 13.04 & 2.78 & 2.436 \\
CP-PAW & 5.693 & 5.666 & 13.008 & 1.98 & 2.396 \\
\hline
\end{tabular}\label{T2BaFeAs}
\end{table}

Fig.~\ref{fig:SUSAsVsSb} presents the comparison of normalized ${\bf
q}$-dependent Pauli susceptibilities at fixed $q_z=\pi$ between
arsenides and antimonides in 111, 1111 and 122 compounds. The
normalization factors are the susceptibilities of the corresponding
arsenide for each type of compounds at ${\bf q}_0=(0,0,\pi)$. We fix
$q_z=\pi$ because of the fact that spins on iron are arranged
antiferromagnetically along the $z$ direction as observed in
experiments~\cite{Cruz08,Huang08,Rotter08PRB,Goldman08,Zhao08}.
However, we checked that the conclusion drawn below will not be
changed if $q_z=0$ is fixed.

\begin{figure}[tb]
\includegraphics[width=0.45\textwidth]{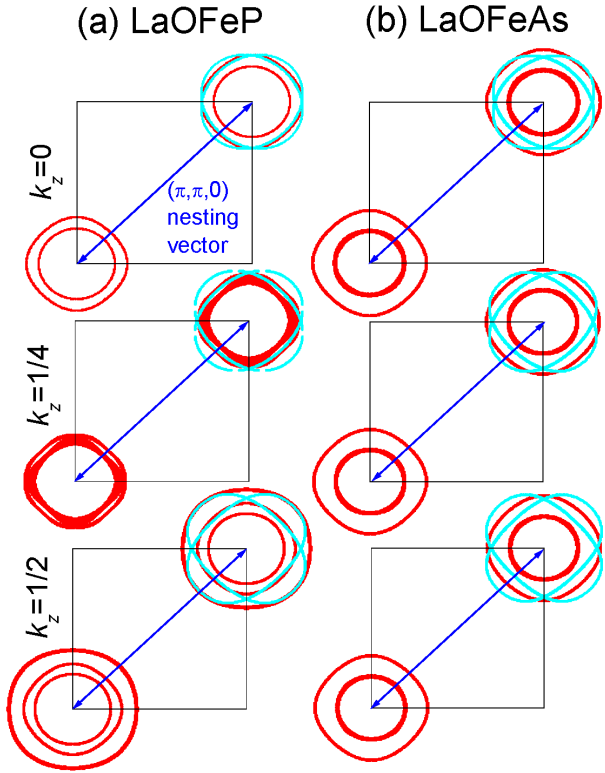}
\caption{(Color online) Fermi surface cuts for (a) LaOFeP and (b)
  LaOFeAs along different $k_z$-planes, where $k_z=n/4$, $n\in
  \{0,1,2\}$ in units of $2\pi$.  The cyan (light gray) curves are the
  electron Fermi surfaces around $(\pi,\pi,k_z)$ and the red (dark
  gray) curves the hole Fermi surfaces around $(0,0,k_z)$. In order to show the
  nesting properties, the hole Fermi surfaces at $(0,0,k_z)$ are shown
  again, shifted by $(\pi,\pi)$. Here we use GGA for the DFT calculations.} \label{fig:FSLaPAs}
\end{figure}

We find that in all Fe-based families, the situation is similar. Two
peaks around ${\bf q}_0=(0,0,\pi)$ and ${\bf q}_{\pi}=(\pi,\pi,\pi)$
are detected in both arsenides and antimonides, and the peaks around
$(\pi,\pi,\pi)$ are always stronger than those around $(0,0,\pi)$,
indicating that the instability towards stripe-type
antiferromagnetic ordering dominates, which is consistent with our
spin-polarized GGA calculations. Most importantly, we find that, in
contrast to Ref.~\onlinecite{Moon09}, the Pauli susceptibilities are
also enhanced together with the magnetizations when As is
substituted by Sb in LaOFeAs, LiFeAs and BaFe$_2$As$_2$, which
demonstrates a connection between Fermi surface nesting and
magnetism and consequently strongly suggests that the theory of
Fermi-surface-nesting-driven magnetism is still valid.

\begin{figure}[tb]
\includegraphics[width=0.45\textwidth]{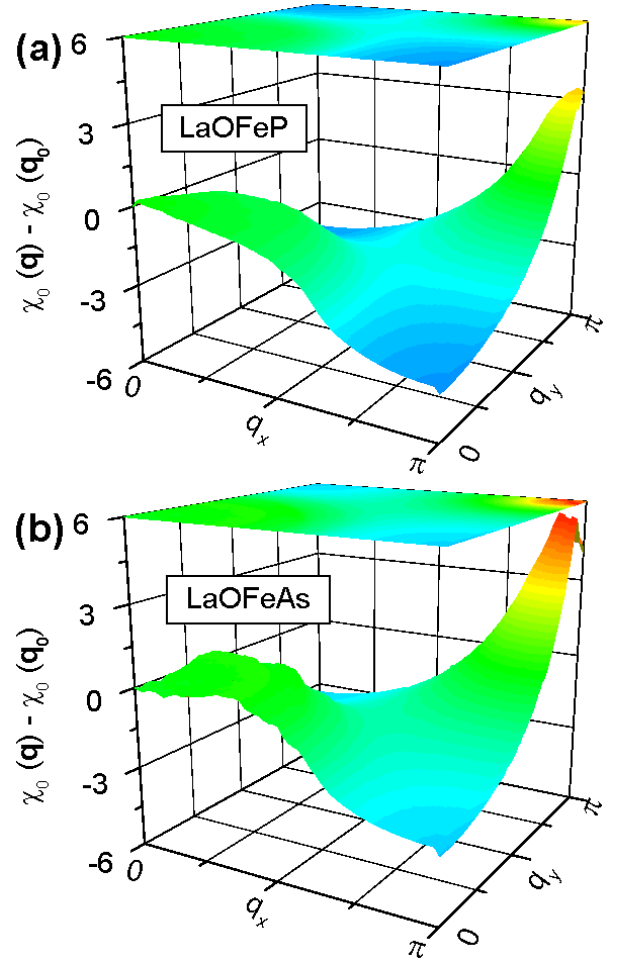}
\caption{(Color online) Static {\bf q}-dependent Pauli
susceptibilities at
  fixed $q_z=\pi$ for (a) LaOFeP and (b) LaOFeAs. The corresponding
  values of the Pauli susceptibilities at ${\bf q}_0=(0,0,\pi)$ in
  LaOFeP and LaOFeAs, respectively, are subtracted. On top,
  two-dimensional contour maps are shown. Here, GGA is used for the DFT calculations.} \label{fig:SUSLaPAs}
\end{figure}

\section{Competition of instabilities in Pauli susceptibilities in 1111 compounds}
\label{sec:four}

\begin{table}[!h]
\caption{Comparison of representative distances $d_{\text{Fe}-Pn}$ in
(\AA) where $Pn$=P, As, Sb and Bi in LaOFeP, LaOFeAs, LaOFeSb, and
LaOFeBi between different GGA optimized structures and experimental
structures, if available.}
\begin{tabular}{ccccccccccccc}
\hline\hline
& Exp.~\cite{McQueen08,Nomura08}  & CP-PAW & VASP~\cite{CSLiu09}  & SIESTA~\cite{Moon09} \\
\hline
LaOFeP & 2.289 & 2.264 & 2.232 & - \\
LaOFeAs & 2.408 & 2.372 & 2.357 & 2.446 \\
LaOFeSb & - & 2.547 & 2.50 & 2.660 \\
LaOFeBi & - & 2.639 & - & - \\
\hline
\end{tabular}\label{T3LaOFePn}
\end{table}

\begin{table}[!h]
\caption{Comparison of representative distances $d_\text{La-O}$ in
(\AA) in LaOFeP, LaOFeAs, LaOFeSb, and LaOFeBi between different
GGA optimized structures and experimental structures, if available.}
\begin{tabular}{ccccccccccccc}
\hline\hline
& Exp.  & CP-PAW (ours) & VASP (ref.) & SIESTA~\cite{Moon09} \\
\hline
LaOFeP & 2.350 & 2.344 & 2.349 & - \\
LaOFeAs & 2.363 & 2.356 & 2.369 & 2.374 \\
LaOFeSb & - & 2.375 & 2.394 & 2.398 \\
LaOFeBi & - & 2.382 & - & - \\
\hline
\end{tabular}\label{T4LaOFePn}
\end{table}

In what follows we concentrate on the 1111 compounds and perform a
comparative study among  LaOFe$Pn$ ($Pn$=P, As, Sb, Bi). In
Table~\ref{T3LaOFePn} and~\ref{T4LaOFePn}, we first present the
comparisons of two representative atomic distances among different structures
optimized within GGA and experimental structures, if available. Our
results agree well with the experimental ones.

Fig.~\ref{fig:FSLaPAs} shows the calculated Fermi surface cuts for
LaOFeP and LaOFeAs on different $k_z$-planes based on the
experimental lattice structures. The figure is almost unchanged if
we consider the optimized lattice structure. The hole Fermi surfaces
around $(0,0,k_z)$ are shifted by $(\pi,\pi,0)$ to show the nesting
properties. Shifting of $(\pi,\pi,\pi)$ was also investigated, and
we find that the nesting properties are nearly unchanged. From the figure,
it is apparent that the Fermi surface nesting is even more perfect
in LaOFeP than in LaOFeAs, indicating a stronger tendency to
stripe-type antiferromagnetic ordering in LaOFeP compared to
LaOFeAs. However, experimentally, while a small magnetization is
observed in undoped LaOFeAs~\cite{Cruz08}, superconductivity rather
than magnetic order is detected in undoped LaOFeP~\cite{Kamihara06}.
These observations would indicate that Fermi surface nesting might
not be connected to magnetization.

\begin{figure}[tb]
\includegraphics[width=0.45\textwidth]{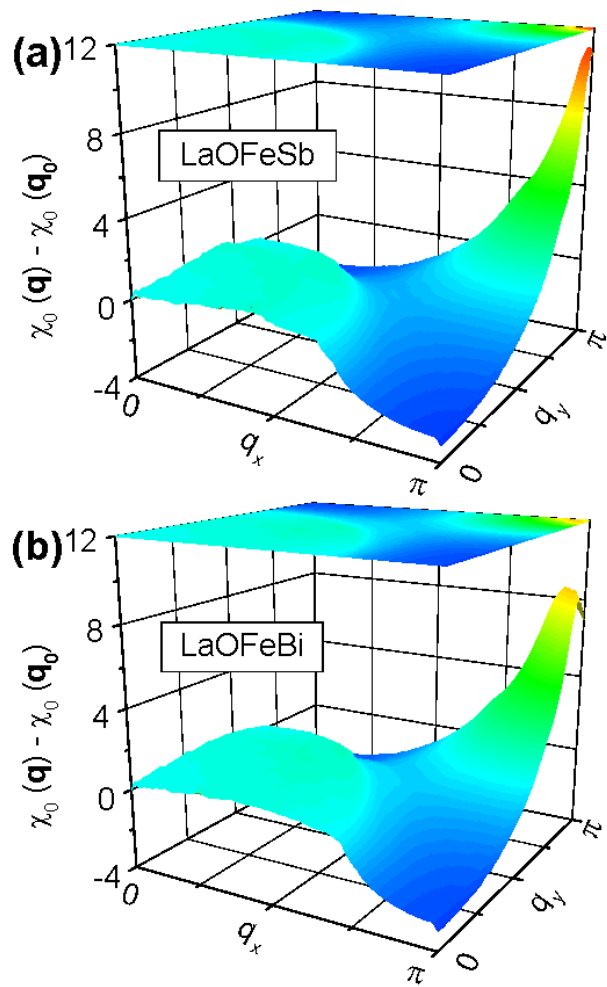}
\caption{(Color online) Static {\bf q}-dependent Pauli
susceptibilities at
  fixed $q_z=\pi$ for (a) LaOFeSb and (b) LaOFeBi. The corresponding
  values of the Pauli susceptibilities at ${\bf q}_0=(0,0,\pi)$ in
  LaOFeSb and LaOFeBi, respectively, are
  subtracted. On top,
  two-dimensional contour maps are shown. In the DFT calculations GGA is used.} \label{fig:SUSLaSbBi}
\end{figure}

In order to quantify the Fermi surface nesting, we show in
Fig.~\ref{fig:SUSLaPAs} the {\bf q}-dependent Pauli susceptibilities
at fixed $q_z=\pi$ for LaOFeP and LaOFeAs with subtraction of the
corresponding values at ${\bf q}_0=(0,0,\pi)$. While a peak in
LaOFeP appears right at ${\bf q}_{\pi}=(\pi,\pi,\pi)$ indicating
almost perfect nesting properties of the Fermi surfaces, peaks are
situated close to ${\bf q}_{\pi}=(\pi,\pi,\pi)$ in LaOFeAs
suggesting nearly nested Fermi surfaces, which is consistent with
the Fermi surface cuts shown in Fig.~\ref{fig:FSLaPAs}. The most
interesting finding in Fig.~\ref{fig:SUSLaPAs} is that the relative
values $\chi(\pi,\pi,\pi)$-$\chi(0,0,\pi)$ increase from LaOFeP to
LaOFeAs irrespective of whether the Fermi surface nesting is perfect
or not. While the peak at $(\pi,\pi,\pi)$ favors stripe-type
antiferromagnetic ordering, the one at $(0,0,\pi)$ represents a
possible instability towards checkerboard-type antiferromagnetic
ordering or A-type antiferromagnetic ordering where ferromagnetic
layers are stacked antiferromagnetically. The heights of these two peaks
become closer in LaOFeP than in LaOFeAs, implying that competition
between the above-mentioned two types of antiferromagnetic states
becomes stronger in LaOFeP if thermal or quantum fluctuations are
taken into account. Also spin-fluctuation mediated pairing of the
superconducting state~\cite{Mazin08,Chubukov08,Graser09} coming from
inter-band scattering around ${\bf q}_{\pi}=(\pi,\pi,\pi)$ takes
part in the competition.  Eventually, as the two types of
antiferromagnetism strongly compete with each other, the additional
superconducting state order emerges and opens a gap, removing the
high instability at the Fermi level and lowering the total energy.
This could be the scenario to explain why undoped LaOFeP is always
nonmagnetic but shows superconductivity below 3.2~K at ambient
pressure. As $\chi(\pi,\pi,\pi)-\chi(0,0,\pi)$ increases beyond a
critical value, the stripe-type antiferromagnetic ordering prevails
over the checkerboard-type one and the pairing.  This scenario may
apply to the low-temperature magnetic phase of LaOFeAs. Furthermore,
we have calculated the total energies of checkerboard and
stripe-type antiferromagnetic phases for both LaOFeP and LaOFeAs. We
found that the stripe-type antiferromagnetic phases are the
ground state in both cases and the energy difference between the two
phases is smaller in LaOFeP than in LaOFeAs, which is consistent
with the trend of $\chi(\pi,\pi,\pi)$-$\chi(0,0,\pi)$.

\begin{figure}[tb]
\includegraphics[angle=-90,width=0.45\textwidth]{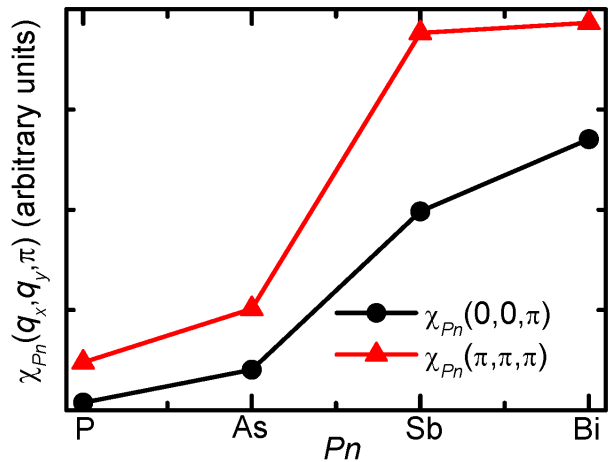}
\caption{(Color online) Static Pauli susceptibilities at ${\bf
q}_0=(0,0,\pi)$
  and ${\bf
q}_{\pi}=(\pi,\pi,\pi)$ for LaOFe$Pn$ with $Pn$=P, As, Sb,
  Bi. Here we use GGA for the DFT calculations.} \label{fig:SusLaPn}
\end{figure}

In Fig.~\ref{fig:SUSLaSbBi}, we display the {\bf q}-dependent Pauli
susceptibilities at fixed $q_z=\pi$ for the hypothetical compounds
LaOFeSb and LaOFeBi. The corresponding values of the Pauli
susceptibilities at ${\bf q}_0=(0,0,\pi)$ in LaOFeSb and LaOFeBi,
respectively, are again subtracted. Note that
$\chi(\pi,\pi,\pi)-\chi(0,0,\pi)$ in LaOFeSb and LaOFeBi is even
larger than that in LaOFeAs (see Fig.~\ref{fig:SUSLaPAs}), indicating
that the instability towards stripe-type antiferromagnetic ordering
could win the competition between different instabilities in these two
compounds. It is also interesting to note that the peak at ${\bf
  q}_0=(0,0,\pi)$ becomes flatter when we go from LaOFeP to
LaOFeBi. While the flatness of the peak can be associated with a
larger number of different magnetic structures lying within a small
energy window, the stronger peaks around ${\bf
q}_{\pi}=(\pi,\pi,\pi)$ compared to ${\bf q}_0=(0,0,\pi)$ in all the
1111 compounds we studied makes other magnetic orderings besides the
stripe-type antiferromagnetic one less favorable. Combining the
results from spin-polarized GGA (LSDA) calculations where magnetic
moments on each iron are given as 2.2~(1.3) and 2.4~(1.8) $\mu_B$ in
LaOFeSb and LaOFeBi, respectively, we predict that the ground states
of these two compounds at ambient pressure without doping should
show stripe-type antiferromagnetic order although spin-polarized GGA
(LSDA) calculations overestimate the magnetic moments.

The increase of the magnetic moment from As to Sb to Bi can be
understood from Fig.~\ref{fig:SusLaPn} where Pauli susceptibilities
at ${\bf q}_0=(0,0,\pi)$ and ${\bf q}_{\pi}=(\pi,\pi,\pi)$ for
LaOFe$Pn$ with $Pn$=P, As, Sb, Bi are explicitly shown. While the
increasing absolute value of $\chi_{Pn}(\pi,\pi,\pi)$ as $Pn$
changes from P to Bi is probably responsible for the increasing
magnetic moment, the difference between $\chi_{Pn}(\pi,\pi,\pi)$ and
$\chi_{Pn}(0,0,\pi)$ dominates the possible competition between
different ordered states, which again implies a strong relation
between Fermi surface nesting and magnetism. Due to the fact that
LaOFeAs is not a superconductor without doping or application of
pressure, we argue that possible superconducting states in LaOFeSb
and LaOFeBi can only occur under doping or application of pressure.

\begin{figure}[tb]
\includegraphics[width=0.45\textwidth]{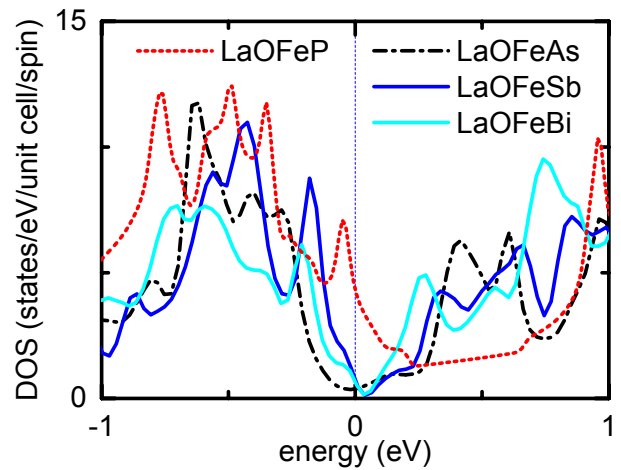}
\caption{(Color online) Total density of states (DOS) for LaOFeAs,
LaOFeSb
  and LaOFeBi in the low-temperature orthorhombic phase with
  stripe-type antiferromagnetic order calculated by spin-polarized
  GGA. Also shown for comparison is the DOS for nonmagnetic LaOFeP calculated by GGA. Relatively high DOS at the Fermi level is present in LaOFeP, indicating
  the possible instability with respect to superconductivity in the absence of magnetic order.}
\label{fig:DOSLaSbBi}
\end{figure}

In Fig.~\ref{fig:DOSLaSbBi}, we display the DOS for LaOFeAs, LaOFeSb
and LaOFeBi calculated within spin-polarized GGA calculations. It is
shown that in all three cases the DOS at the Fermi level remains
finite, suggesting that undoped LaOFeSb and LaOFeBi are stripe-type
antiferromagnetic metals at ambient pressure as is the case of
LaOFeAs or other iron-based superconductors. Our results from
spin-polarized GGA calculations cannot corroborate the arguments
introduced in Ref.~\onlinecite{Baskaran08} where it is argued that
LaOFeSb and LaOFeBi could be antiferromagnetic Mott insulators,
though the tendency towards larger Fe magnetic moments from the As
to Bi compounds is observed. Such a conclusion is also confirmed by
our LSDA calculations. In Fig.~\ref{fig:DOSLaSbBi}, the DOS for
LaOFeP is also shown. Since LaOFeP is nonmagnetic at low
temperature, only the non-spinpolarized GGA result is presented.
It is found that the DOS
at the Fermi level remains relatively high, which indicates a possible
instability with respect to superconductivity in the absence of magnetic
order with the opening of a superconducting gap and the lifting of the
degeneracy at the Fermi level.

\section{ New superconductor candidates}
\label{sec:five}

\begin{figure}[tb]
\includegraphics[angle=-90,width=0.45\textwidth]{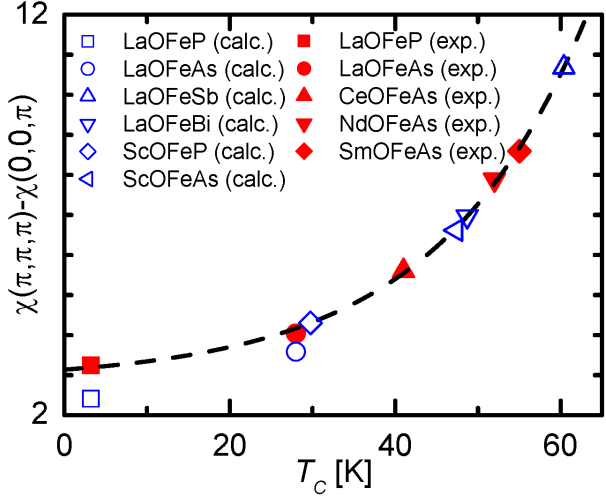}
\caption{(Color online) Prediction of superconducting transition
  temperatures $T_c$ from a phenomenological relation between $T_c$
  and $\chi(\pi,\pi,\pi)-\chi(0,0,\pi)$ for the parent compounds of
  the hypothetical 1111 compounds LaOFeSb, LaOFeBi, ScOFeP and
  ScOFeAs. The phenomenological relation was determined by
  calculating $\chi(\pi,\pi,\pi)- \chi(0,0,\pi)$ for several 1111
  compounds LaOFeP, LaOFeAs, CeOFeAs, NdOFeAs, and SmOFeAs by GGA where $T_c$'s and lattice structures are given
  experimentally. The $\chi(\pi,\pi,\pi)- \chi(0,0,\pi)$ for LaOFeP and LaOFeAs calculated by GGA from DFT optimized structures are also shown for comparison. It is found that the
resulting $\chi(\pi,\pi,\pi)-\chi(0,0,\pi)$ for LaOFeP and LaOFeAs
based on our optimized structures is only slightly underestimated
compared to that calculated from experimental structures, which
shows that the result depends only weakly on our structure
optimization. } \label{fig:TcPredictSUS}
\end{figure}

Experimentally, the highest recorded superconducting transition
temperature has been observed in the 1111 compounds. It is therefore
tempting to find ways to predict $T_c$ for hypothetical 1111
compounds. One promising route is to consider a phenomenological
relation between $T_c$ and the $\chi(\pi,\pi,\pi)-\chi(0,0,\pi)$ of
the parent compound. As we know, with doping or application of
pressure, the $\chi(\pi,\pi,\pi)$ instability is suppressed,
resulting in the disappearance of stripe-type antiferromagnetic
order. However, strong inter-band scattering with a wave vector
around $(\pi,\pi,\pi)$ compared to intra-band scattering with a wave
vector around $(0,0,\pi)$ remains, which leads to a superconducting
state. Therefore we argue that the larger the relative value of
$\chi(\pi,\pi,\pi)$-$\chi(0,0,\pi)$ in the parent compound is, the
stronger the antiferromagnetic spin fluctuation after suppression of
magnetic order, and thus the higher the superconducting transition
temperature will be.

In Fig.~\ref{fig:TcPredictSUS}, we plot
$\chi(\pi,\pi,\pi)-\chi(0,0,\pi)$ versus $T_c$ for the hypothetical
1111 compounds LaOFeSb, LaOFeBi, ScOFeP and ScOFeAs.  The
phenomenological relation between $T_c$ and
$\chi(\pi,\pi,\pi)-\chi(0,0,\pi)$ is determined by first calculating
$\chi(\pi,\pi,\pi)- \chi(0,0,\pi)$ for several typical 1111
compounds LaOFeP~\cite{Kamihara06,McQueen08},
LaOFeAs~\cite{Kamihara08,Nomura08},
CeOFeAs~\cite{Chen08Ce,Zhao08Ce}, NdOFeAs~\cite{Ren08Nd,Qiu08Nd},
and SmOFeAs~\cite{Ren08Sm,Martinelli08Sm}, where $T_c$ and lattice
structures are given experimentally, and then fitting the data by an
exponential growth function of
$\chi(\pi,\pi,\pi)-\chi(0,0,\pi)=2.86+0.28*\exp(T_c/18.14)$. With
this relation, Tc for the compounds which have not yet been
experimentally reported is predicted by optimizing the lattice
structure and calculating $\chi(\pi,\pi,\pi)-\chi(0,0,\pi)$ from DFT
calculations. From Fig.~\ref{fig:TcPredictSUS}, we find that the
resulting $\chi(\pi,\pi,\pi)-\chi(0,0,\pi)$ for LaOFeP and LaOFeAs
based on our optimized structures is only slightly underestimated
compared to that calculated from experimental structures, which
shows that the result depends only weakly on our structure
optimization. Among the four 1111 compounds, this procedure shows
that LaOFeSb can give the highest $T_c$ around 60~K which is above
the highest recorded $T_c$ of 55~K in SmOFeAs.

\begin{figure}[tb]
\includegraphics[angle=-90,width=0.45\textwidth]{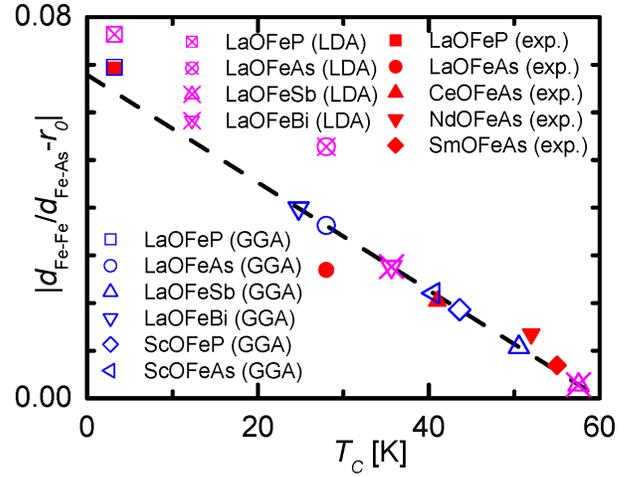}
\caption{(Color online) Prediction of superconducting transition
  temperatures $T_c$ from a phenomenological relation between $T_c$
  and the ratio of atomic distances $d_\text{Fe-Fe}$ and
  $d_\text{Fe-As}$ for the hypothetical 1111 compounds LaOFeSb, LaOFeBi, ScOFeP and
  ScOFeAs. A constant ratio of
  $r_0=d_{0,\text{Fe-Fe}}/d_{0,\text{Fe-As}}$ for distances
  $d_{0,\text{Fe-Fe}}$, $d_{0,\text{Fe-As}}$ in a perfect FeAs$_4$
  tetrahedron is subtracted. The phenomenological relation is determined by taking into
account the same compounds LaOFeP, LaOFeAs, CeOFeAs, NdOFeAs, and
SmOFeAs as in Fig.~\ref{fig:TcPredictSUS} where $T_c$ and lattice
structures are given experimentally and doing a linear fit as in
Ref.~\onlinecite{Zhao08Ce}. The
$|d_\text{Fe-Fe}/d_\text{Fe-As}-r_0|$ for LaOFeP, LaOFeAs, LaOFeSb,
and LaOFeBi based on optimized structures within GGA and LDA are
also shown for comparisons. Comparing the results for LaOFeP and
LaOFeAs from GGA and LDA optimizations with those from experiments,
we find that the optimized structures from GGA are more consistent
with the experimental one.} \label{fig:TcPredict}
\end{figure}

An alternative procedure to predict $T_c$ phenomenologically is
based on the fact that the physical properties of Fe-based
superconductors strongly depend on the As position. Tiny shiftings
of the As position away from or closer to the iron plane will
significantly change the band structure around the Fermi
level~\cite{Singh08,Vildosola08}. Therefore, similar to
Refs.~\onlinecite{Zhao08Ce} and~\onlinecite{Lee08Struct}, we plot in
Fig.~\ref{fig:TcPredict} $T_c$ versus the absolute value of the
ratio between atomic distances $d_\text{Fe-Fe}$ and $d_\text{Fe-As}$
while subtracting $r_0=d_{0,\text{Fe-Fe}}/d_{0,\text{Fe-As}}$ where
d$_{0,\text{Fe-Fe}}$ and d$_{0,\text{Fe-As}}$ denote the distances
in a perfect tetrahedron formed by four nearest neighbor As atoms
surrounding one Fe atom. Similar to the first scheme, we determine
the phenomenological relation between
$|d_\text{Fe-Fe}/d_\text{Fe-As}-r_0|$ and $T_c$ by taking into
account the same compounds LaOFeP, LaOFeAs, CeOFeAs, NdOFeAs, and
SmOFeAs as in the first scheme where $T_c$ and lattice structures
are given experimentally and doing a linear fit as in
Ref.~\onlinecite{Zhao08Ce}. With this relation, the $T_c$ for
LaOFeSb, LaOFeBi, ScOFeP and ScOFeAs is predicted based on the
optimized structure.

Fig.~\ref{fig:TcPredict} presents the results based on the
structures optimized within both GGA and LDA. Comparing the results
for LaOFeP and LaOFeAs from GGA and LDA optimizations with those
from experiments, we find that the optimized structures from GGA are
more consistent with the experimental one. According to the relation
we fitted, LaOFeSb always gives the highest $T_c$ of 57.5~(50.6)~K
for LDA (GGA) optimizations, respectively, among the four 1111
compounds we studied. Combining the two presented phenomenological
prediction schemes, we clearly obtain that LaOFeSb would be upon
doping or under pressure a good candidate for superconductivity with
highest $T_c$ and it would be very interesting to see it
synthesized.

\section{ Conclusions}
\label{sec:six}

In the present work we studied the physical properties of LaOFe$Pn$,
BaFe$_2Pn_2$ and LiFe$Pn$ with $Pn$=As and Sb. Our results support
the validity of the itinerant nature of magnetism in these compounds
where magnetization is closely related to Fermi surface nesting.
Furthermore, we concentrated on the 1111 compounds LaOFe$Pn$ with
$Pn$=P, As, Sb, and Bi.  We found that the increase of the magnetic
moment in the undoped compounds with $Pn$ varying from P to Bi is
due to the increasing instability of the Pauli susceptibility at
${\bf q}_{\pi}=(\pi,\pi,\pi)$ and the decreasing competition to the
instability at ${\bf
  q}_0=(0,0,\pi)$. The superconducting state appearing in undoped
LaOFeP at ambient pressure is ascribed to the strong competition
between the instability at ${\bf q}_{\pi}=(\pi,\pi,\pi)$ and ${\bf
  q}_0=(0,0,\pi)$. Thus, together with the investigation of the DOS in
the low temperature phase, we argue that the hypothetical compounds
LaOFeSb and LaOFeBi are antiferromagnetic metals at ambient pressure
without doping. The results for LaOFe$Pn$ again strongly imply that
Fermi surface nesting plays a dominating role in the physical
properties of the 1111 compounds. Finally we consider two
phenomenological relations to predict the superconducting transition
temperature $T_c$ for the hypothetical 1111 compounds and predict
that LaOFeSb would be a possible candidate for a superconductor with
a higher $T_c$ than presently recorded for the known Fe-based
superconductors.  Combining the fact that Fermi surface nesting
dominates the physics in the 122 compounds, 1111 compounds, 11
compounds, and 111 compounds, we argue that magnetism in iron-based
superconductors is strongly influenced by their itinerant nature.
However, from our study, the localized scenario is not ruled out and
may also play an important role in the physics of iron-based
superconductors. Furthermore, while in our study we emphasize the
role of the states at the Fermi level and accordingly the nesting
property of the Fermi surface on the itinerant nature of magnetism, the
significant contributions to the finite moment of itinerant
magnetism from the states in the vicinity of the Fermi level should not be
ignored.

{\it Acknowledgments.-} We would like to thank the Deutsche
Forschungsgemeinschaft for financial support through the SFB/TRR~49
and Emmy Noether programs.


\begin{thebibliography}{99}
\bibitem{Kamihara08} Y. Kamihara, T. Watanabe, M. Hirano, and
  H. Hosono, J.  Am. Chem. Soc. {\bf 130}, 3296 (2008).

\bibitem{Ren08} Z. A. Ren, W. Lu, J. Yang, W. Yi, X. L. Shen,
  Z. C. Li, G. C. Che, X. L. Dong, L. L. Sun, F. Zhou, and Z. X. Zhao,
  Chin. Phys. Lett. {\bf 25}, 2215 (2008).

\bibitem{Torikachvili08} Milton S. Torikachvili, Sergey L. Bud'ko, Ni
  Ni, and Paul C.  Canfield, Phys. Rev. Lett. {\bf 101}, 057006
  (2008).

\bibitem{Sasmal08} K. Sasmal, B. Lv, B. Lorenz, A. M. Guloy, F. Chen,
  Y. Xue and P. C.  W. Chu, Phys. Rev. Lett. {\bf 101}, 107007 (2008).

\bibitem{Chen08} G. F. Chen, Z. Li, G. Li, Z. Hu, J. Dong,
  X. D. Zhang, P. Zheng, N.  L. Wang and J. L. Luo,
  Chin. Phys. Lett. {\bf 25}, 3403 (2008).

\bibitem{Rotter08PRL} M. Rotter, M. Tegel and D. Johrendt,
  Phys. Rev. Lett. {\bf 101}, 107006 (2008).

\bibitem{Alireza09} P. L. Alireza, Y. T. C. Ko, J. Gillett,
  C. M. Petrone, J. M. Cole, G. G. Lonzarich and S. E. Sebastian,
  J. Phys.: Condens. Matter {\bf 21}, 012208 (2009).

\bibitem{Kimber09} S. A. J. Kimber, A. Kreyssig, Y.-Z. Zhang, H. O.
  Jeschke, R. Valent{\'\i}, F. Yokaichiya, E. Colombier, J. Yan,
  T. C. Hansen, T. Chatterji, R. J.  McQueeney, P. C. Canfield,
  A. I. Goldman and D. N. Argyriou, Nature Materials {\bf 8}, 471
  (2009).

\bibitem{Sefat08} A. S. Sefat, R. Jin, M. A. McGuire, B. C. Sales,
  D. J. Singh, and D.  Mandrus, Phys. Rev. Lett. {\bf 101}, 117004
  (2008).

\bibitem{Wang08} X. C. Wang, Q. Liu, Y. Lv, W. Gao, L. X. Yang,
  R. C. Yu, F. Y. Li, and C. Jin, Solid. State. Commun. {\bf 148}, 538
  (2008)

\bibitem{Tapp08} J. H. Tapp, Z. Tang, B. Lv, K. Sasmal, B. Lorenz,
  P. C. W. Chu, and A. M. Guloy, Phys. Rev. B {\bf 78}, 060505 (2008).

\bibitem{Pitcher08} M. J. Pitcher, D. R. Parker, P. Adamson,
  S. J. C. Herkelrath, A. T.  Boothroyd, R. M. Ibberson, M. Brunelli,
  and S. J. Clarke, Chem.  Commun. {\bf 2008}, 5918 (2008).

\bibitem{SJZhang09} S. J. Zhang, X. C. Wang, R. Sammynaiken,
  J. S. Tse, L. X. Yang, Z.  Li, Q. Q. Liu, S. Desgreniers, Y. Yao,
  H. Z. Liu, and C. Q. Jin, Phys. Rev. B {\bf 80}, 014506 (2009).

\bibitem{Parker09} D. R. Parker, M. J. Pitcher, P. J. Baker,
  I. Franke, T. Lancaster, S. J. Blundell, and S. J. Clarke,
  Chem. Commun. (Cambridge) ({\bf 2009}), 2189.

\bibitem{Chen09} G. F. Chen, W. Z. Hu, J. L. Luo, and N. L. Wang,
  Phys. Rev. Lett.  102, 227004 (2009).

\bibitem{Hsu08} F.-C. Hsu, J.-Y. Luo, K.-W. Yeh, T.-K. Chen,
  T.-W. Huang, P. M. Wu, Y.-C. Lee, Y.-L. Huang, Y.-Y. Chu, D.-C. Yan,
  and M.-K. Wu, Proc.  Natl. Acad. Sci. U.S.A. {\bf 105}, 14262
  (2008).

\bibitem{Mizuguchi08} Y. Mizuguchi, F. Tomioka, S. Tsuda,
  T. Yamaguchi, and Y. Takano, Appl. Phys. Lett. {\bf 93}, 152505
  (2008).

\bibitem{Yeh08} K.-W. Yeh, T.-W. Huang, Y. lin Huang, T.-K. Chen,
  F.-C. Hsu, P. M.  Wu, Y.-C. Lee, Y.-Y. Chu, C.-L. Chen, J.-Y. Luo,
  D.-C. Yan, and M.-K. Wu, Europhys. Lett. {\bf 84} 37002 (2008).

\bibitem{Yuan09} H. Q. Yuan, J. Singleton, F. F. Balakirev, S. A. Baily, G. F. Chen, J. L. Luo, N. L. Wang, Nature {\bf 457}, 565 (2009).

\bibitem{Singh08} D. J. Singh and M.-H. Du, Phys. Rev. Lett. {\bf
  100}, 237003 (2008).

\bibitem{Haule08} K. Haule, J. H. Shim, and G. Kotliar,
  Phys. Rev. Lett. {\bf 100}, 226402 (2008).

\bibitem{Boeri08} L. Boeri, O. V. Dolgov, and A. A. Golubov,
  Phys. Rev. Lett. {\bf 101}, 026403 (2008).

\bibitem{Mazin08} I. I. Mazin, D. J. Singh, M. D. Johannes and
  M. H. Du, Phys. Rev. Lett. {\bf 101}, 057003 (2008).

\bibitem{Mazin09} I. I. Mazin, J. Schmalian, Physica C {\bf 469}, 614
  (2009).

\bibitem{Wang09} F. Wang, H. Zhai, and D. Lee, Europhys. Lett. {\bf
  85}, 37005 (2009).

\bibitem{Chubukov08} A. V. Chubukov, D. Efremov, and I. Eremin,
  Phys. Rev. B {\bf 78}, 134512 (2008).

\bibitem{Stanev08} V. Stanev, J. Kang, and Z. Tesanovic, Phys. Rev. B
  {\bf 78}, 184509 (2008).

\bibitem{Kuroki08} K. Kuroki, S. Onari, R. Arita, H. Usui, Y. Tanaka,
  H.  Kontani, and H. Aoki, Phys. Rev. Lett. {\bf 101}, 087004 (2008).

\bibitem{Graser08} S. Graser, G. R. Boyd, C. Cao, H.-P. Cheng, P. J.
  Hirschfeld, and D. J. Scalapino, Phys. Rev. B {\bf 77}, 180514(R)
  (2008).

\bibitem{Graser09} S. Graser, T. A. Maier, P. J. Hirschfeld, D. J.
  Scalapino, New J. Phys. {\bf 11}, 025016 (2009).

\bibitem{Qi08} X.-L. Qi, S. Raghu, C.-X. Liu, D. J. Scalapino, and
  S.-C. Zhang, arXiv:0804.4332 (unpublished).

\bibitem{Yao09} Z.-J. Yao, J.-X. Li, and Z. D. Wang, New J. Phys. {\bf
  11}, 025009 (2009).

\bibitem{Sknepnek09} R. Sknepnek, G. Samolyuk, Y. Lee, J. Schmalian,
  Phys. Rev. B {\bf 79}, 054511 (2009).

\bibitem{Si08} Q. Si and E. Abrahams, Phys. Rev. Lett. {\bf 101},
  076401 (2008).

\bibitem{Yildirim08} T. Yildirim, Phys. Rev. Lett. {\bf 101}, 057010
  (2008).

\bibitem{Seo08} K. Seo, B. A. Bernevig, and J. Hu, Phys. Rev. Lett.
  {\bf 101}, 206404 (2008).

\bibitem{Ishida09} K. Ishida, Y. Nakai, H. Hosono, J. Phys. Soc. Jpn.
  {\bf 78}, 062001 (2009).

\bibitem{Hsieh08} D. Hsieh, Y. Xia, L. Wray, D. Qian, K. Gomes,
  A. Yazdani, G. F. Chen, J. L. Luo, N. L. Wang, M. Z. Hasan,
  arXiv:0812.2289 (unpublished).

\bibitem{Fink09} J. Fink, S. Thirupathaiah, R. Ovsyannikov,
  H. A. D\"urr, R. Follath, Y. Huang, S. de Jong, M. S. Golden,
  Yu-Zhong Zhang, H. O. Jeschke, R. Valent{\'\i}, C. Felser,
  S. Dastjani Farahani, M. Rotter, and D.  Johrendt, Phys. Rev. B {\bf
    79}, 155118 (2009).

\bibitem{Cvetkovic09} V. Cvetkovic and Z. Tesanovic,
  Europhys. Lett. {\bf 85}, 37002 (2009).

\bibitem{Dong08} J. Dong, H. J. Zhang, G. Xu, Z. Li, G. Li, W. Z. Hu,
  D. Wu, G. F.  Chen, X. Dai, J. L. Luo, Z. Fang, and N. L. Wang,
  Europhys. Lett. {\bf 83}, 27006 (2008).

\bibitem{Opahle08} I. Opahle, H. C. Kandpal, Y. Zhang, C. Gros, and R.
  Valent{\'\i}, Phys. Rev. B {\bf 79}, 024509 (2009).

\bibitem{YZZhang09} Y.-Z. Zhang, H. C. Kandpal, I. Opahle,
  H. O. Jeschke, and R.  Valent{\'\i}, Phys. Rev. B {\bf 80}, 094530
  (2009).

\bibitem{Yaresko09} A. N. Yaresko, G.-Q. Liu, V. N. Antonov, O.K. Andersen, Phys. Rev. B {\bf 79}, 144421
  (2009).

\bibitem{Kroll08} T. Kroll, S. Bonhommeau, T. Kachel, H. A. D\"urr,
  J. Werner, G.  Behr, A. Koitzsch, R. H\"ubel, S. Leger,
  R. Sch\"onfelder, A. K.  Ariffin, R. Manzke, F. M. F. de Groot,
  J. Fink, H. Eschrig, B.  B{\"u}chner, and M. Knupfer, Phys. Rev. B
  {\bf 78}, 220502(R) (2008).

\bibitem{Skornyakov09} S. L. Skornyakov, A. V. Efremov,
  N. A. Skorikov, M. A. Korotin, Yu.  A. Izyumov, V. I. Anisimov,
  A. V. Kozhevnikov, and D. Vollhardt, Phys. Rev. B {\bf 80}, 092501
  (2009).

\bibitem{Yang09} W. L. Yang, A. P. Sorini, C-C. Chen, B. Moritz,
  W.-S. Lee, F.  Vernay, P. Olalde-Velasco, J. D. Denlinger,
  B. Delley, J.-H. Chu, J.  G. Analytis, I. R. Fisher, Z. A. Ren,
  J. Yang, W. Lu, Z. X. Zhao, J.  van den Brink, Z. Hussain,
  Z.-X. Shen, and T. P. Devereaux, Phys.  Rev. B {\bf 80}, 014508
  (2009).

\bibitem{Aichhorn09} M. Aichhorn, L. Pourovskii, V. Vildosola, M.
  Ferrero, O. Parcollet, T. Miyake, A. Georges, and S. Biermann, Phys.
  Rev. B {\bf 80}, 085101 (2009).

\bibitem{Han09A} M. J. Han, Q. Yin, W. E. Pickett, and S. Y. Savrasov,
  Phys. Rev.  Lett. {\bf 102}, 107003 (2009).

\bibitem{Craco08} L. Craco, M. S. Laad, S. Leoni, and H. Rosner,
  Phys. Rev. B {\bf 78}, 134511 (2008).

\bibitem{Ma08} F. Ma, Z.-Y. Lu, T. Xiang, Phys. Rev. B {\bf 78},
  224517 (2008).

\bibitem{Ma09} F. Ma, W. Ji, J. Hu, Z.-Y. Lu, T. Xiang,
  Phys. Rev. Lett. {\bf 102}, 177003 (2009).

\bibitem{Krueger09} F. Kr\"uger, S. Kumar, J. Zaanen, and J. van den Brink, Phys. Rev. B {\bf 79}, 054504 (2009).

\bibitem{LYang09} L. Yang Y. Zhang, H. Ou, J. Zhao, D. Shen, B. Zhou,
  J. Wei, F. Chen, M. Xu, C. He, Y. Chen, Z. Wang, X. Wang, T. Wu,
  G. Wu, X. Chen, M.  Arita, K. Shimada, M. Taniguchi, Z. Lu,
  T. Xiang, and D. Feng, Phys.  Rev. Lett. {\bf 102}, 107002 (2009).

\bibitem{Schmidt09} B. Schmidt, M. Siahatgar, P. Thalmeier,
  arXiv:0911.5664 (unpublished).

\bibitem{Johannes09} M. D. Johannes and I. I. Mazin, Phys. Rev. B {\bf
  79}, 220510(R) (2009).

\bibitem{JWu08} J. Wu, P. Phillips, and A. H. Castro Neto,
  Phys. Rev. Lett. {\bf 101}, 126401 (2008).

\bibitem{JWu09} J. Wu and P. Phillips, arXiv:0901.3538 (unpublished).

\bibitem{Kou08} S.-P. Kou, T. Li, and Z.-Y. Weng, arXiv:0811.4111
  (unpublished).

\bibitem{FWang09} F. Wang, H. Zhai and D.-H. Lee, Europhys. Lett. {\bf
  85}, 37005 (2009).

\bibitem{Barriga09} J. Sanchez-Barriga, J. Fink, V. Boni, I. Di Marco,
  J. Braun, J.  Minar, A. Varykhalov, O. Rader, V. Bellini, F. Manghi,
  H. Ebert, M. I. Katsnelson, A. I. Lichtenstein, O. Eriksson,
  W. Eberhardt, H.  A. D{\"u}rr, Phys. Rev. Lett. {\bf 103}, 267203 (2009).

\bibitem{Cricchio09} F. Cricchio, O. Gr{\aa}n{\"a}s, L. Nordstr{\"o}m,
  arXiv:0911.1342 (unpublished).

\bibitem{Lee09} H. Lee, Y.-Z. Zhang, H. O. Jeschke, and
  R. Valent{\'\i}, arXiv:0912.4024 (unpublished).

\bibitem{Xia09} Y. Xia, D. Qian, L. Wray, D. Hsieh, G. F. Chen,
  J. L. Luo, N. L. Wang, and M. Z. Hasan, Phys. Rev. Lett. {\bf 103},
  037002 (2009).

\bibitem{Moon09} C.-Y. Moon, S. Y. Park, and H. J. Choi, Phys. Rev. B
  {\bf 80}, 054522 (2009).

\bibitem{Li09} S. Li, C. de la Cruz, Q. Huang, Y. Chen, J. W. Lynn,
  J. Hu, Y.-L. Huang, F.-C. Hsu, K.-W. Yeh, M.-K. Wu, and P. Dai,
  Phys. Rev. B {\bf 79}, 054503 (2009).

\bibitem{Han09B} M. J. Han and S. Y. Savrasov, Phys. Rev. Lett. {\bf
  103}, 067001 (2009).

\bibitem{CarParrinello} R. Car, M. Parrinello, Phys. Rev. Lett. {\bf
  55}, 2471 (1985).

\bibitem{Bloechl} P. E. Bl{\"o}chl, Phys. Rev. B {\bf 50}, 17953
  (1994).

\bibitem{Mazin08PRB} I. I. Mazin, M. D. Johannes, L. Boeri,
  K. Koepernik, and D. J.  Singh, Phys. Rev. B {\bf 78}, 085104
  (2008).

\bibitem{Blaha} P. Blaha, K. Schwarz, G. Madsen, D. Kvaniscka, and J. Luitz, WIEN2K, An Augmented Plane
  Wave+Local Orbitals Program for Calculating Crystal, edited by
  K. Schwarz (Techn. University,Vienna, Austria, 2001).

\bibitem{FPLO} K. Koepernik and H. Eschrig, Phys. Rev. B {\bf 59}, 1743
  (1999).  http://www.FPLO.de

\bibitem{Eschrig04} H. Eschrig, M. Richter, and I. Opahle,
in: {\em Relativistic Electronic Structure Theory -
Part II: Applications}, Ed. P. Schwerdtfeger (Elsevier, Amsterdam 2004),
pp. 723--776.

\bibitem{Kozlova05} N. Kozlova, J. Hagel, M. Doerr, J. Wosnitza, D. Eckert,
K.-H. M\"uller, L. Schultz, I. Opahle, S. Elgazzar, Manuel Richter,
G. Goll, H. v. L\"ohneysen, G. Zwicknagl, T. Yoshino, T. Takabatake,
Phys. Rev. Lett. {\bf 95}, 086403 (2005).

\bibitem{Wosnitza06} J. Wosnitza, G. Goll, A. D. Bianchi, B. Bergk,
N. Kozlova, I. Opahle, S. Elgazzar, Manuel Richter, O. Stockert,
H. v. L\"ohneysen, T. Yoshino, and T. Takabatake,
New J. Phys. {\bf 8}, 174 (2006).

\bibitem{GGA} It is well-known that the magnetic moments are
  overestimated by GGA~\cite{YZZhang09,Opahle08,Mazin08PRB} when
  compared to experimental
  data~\cite{SJZhang09,Cruz08,Huang08,Rotter08PRB,Goldman08,Zhao08} or
  even to LDA
  calculations~\cite{Opahle08,Singh08PRB,LZhang08PRB,Kasinathan09}.
  The reason for choosing GGA for the structure optimization is that
  the GGA optimized structures are more consistent with experiment
  than the LDA ones~\cite{YZZhang09,Mazin08PRB}.

\bibitem{Cruz08} Clarina de la Cruz, Q. Huang, J. W. Lynn, Jiying Li,
  W. Ratcliff II, J. L. Zarestky, H. A. Mook, G. F. Chen, J. L. Luo,
  N. L. Wang, and Pengcheng Dai, Nature {\bf 453}, 899 (2008).

\bibitem{Huang08} Q. Huang, Y. Qiu, Wei Bao, M. A. Green, J. W. Lynn,
  Y. C.  Gasparovic, T. Wu, G. Wu, and X. H. Chen,
  Phys. Rev. Lett. {\bf 101}, 257003 (2008).

\bibitem{Rotter08PRB} M. Rotter, M. Tegel, D. Johrendt,
  I. Schellenberg, W. Hermes, and R. P\"ottgen, Phys. Rev. B {\bf 78},
  020503(R) (2008).

\bibitem{Goldman08} A. I. Goldman, D. N. Argyriou, B. Ouladdiaf,
  T. Chatterji, A.  Kreyssig, S. Nandi, N. Ni, S. L. Bud'ko,
  P. C. Canfield, and R. J.  McQueeney, Phys. Rev. B {\bf 78},
  100506(R) (2008).

\bibitem{Zhao08} J. Zhao, W. Ratcliff, II, J. W. Lynn, G. F. Chen,
  J. L. Luo, N. L.  Wang, J. Hu, and P. Dai, Phys. Rev. B {\bf 78},
  140504(R) (2008).

\bibitem{Singh08PRB} D. J.  Singh, Phys. Rev. B {\bf 78}, 094511
  (2008).

\bibitem{LZhang08PRB} L. Zhang, A. Subedi, D. J. Singh, and M. H. Du,
  Phys. Rev. B {\bf 78}, 174520 (2008).

\bibitem{Kasinathan09} D. Kasinathan, A. Ormeci, K. Koch,
  U. Burkhardt, W. Schnelle, A. Leithe-Jasper and H.  Rosner, New J.
  Phys. {\bf 11}, 025023 (2009).

\bibitem{ZLi09} Z. Li, J. S. Tse, and C. Q. Jin, Phys. Rev. B {\bf
  80}, 092503 (2009).

\bibitem{YFLi09} Y.-F. Li, B.-G. Liu, Eur. Phys. J. B {\bf 72}, 153
  (2009).

\bibitem{Moon08} C.-Y. Moon, S. Y. Park, and H. J. Choi, Phys. Rev. B
  {\bf 78}, 212507 (2008).

\bibitem{Vildosola08} V. Vildosola, L. Pourovskii, R. Arita,
  S. Biermann, and A. Georges, Phys. Rev. B {\bf 78}, 064518 (2008).

\bibitem{CSLiu09} C.S. Liu, Y.L. Li, Y. Xu,
  X.L. Wang, and Z. Zeng, Physica B {\bf 404}, 3242 (2009).

\bibitem{Kamihara06} Y. Kamihara, H. Hiramatsu, M. Hirano,
  R. Kawamura, H. Yanagi, T.  Kamiya, and H. Hosono,
  J. Am. Chem. Soc. {\bf 128}, 10012 (2006).

\bibitem{Baskaran08} G. Baskaran, J. Phys. Soc. Jpn. {\bf 77}, 113713
  (2008).

\bibitem{McQueen08} T. M. McQueen, M. Regulacio, A. J. Williams,
  Q. Huang, J. W. Lynn, Y. S. Hor, D. V. West, M. A. Green,
  R. J. Cava, Phys. Rev. B {\bf 78}, 024521 (2008).

\bibitem{Nomura08} T. Nomura, S. W. Kim, Y. Kamihara, M. Hirano,
  P. V. Sushko, K. Kato, M. Takata, A. L. Shluger, H. Hosono,
  Supercond. Sci. Technol. {\bf 21}, 125028 (2008).

\bibitem{Chen08Ce} G. F. Chen, Z. Li, D. Wu, G. Li, W. Z. Hu, J. Dong,
  P. Zheng, J. L.  Luo, and N. L. Wang, Phys. Rev. Lett. {\bf 100},
  247002 (2008).

\bibitem{Zhao08Ce} J. Zhao, Q. Huang, C. de la Cruz, S.  Li,
  J. W. Lynn, Y.  Chen, M. A. Green, G. F. Chen, G. Li, Z. Li,
  J. L. Luo, N. L. Wang, P. Dai, Nature Materials {\bf 7}, 953 (2008).

\bibitem{Ren08Nd} Z.-A. Ren, J. Yang, W. Lu, W. Yi, X.-L. Shen,
  Z.-C. Li, G.-C. Che, X.-L. Dong, L.-L. Sun, F. Zhou, and Z.-X. Zhao,
  Europhys. Lett. {\bf 82}, 57002 (2008).

\bibitem{Qiu08Nd} Y. Qiu, Wei Bao, Q. Huang, T. Yildirim, J. Simmons,
  J. W. Lynn, Y. C.  Gasparovic, J. Li, M. Green, T. Wu, G. Wu,
  X. H. Chen, Phys. Rev.  Lett. {\bf 101}, 257002 (2008).

\bibitem{Ren08Sm} Z.-A. Ren, W. Lu, J. Yang, W. Yi, X.-L. Shen,
  Z.-C. Li, G.-C. Che, X.-L. Dong, L.-L. Sun, F. Zhou, and Z.-X. Zhao,
  Chin. Phys. Lett.  {\bf 25}, 2215 (2008).

\bibitem{Martinelli08Sm} A. Martinelli, M. Ferretti, P. Manfrinetti,
  A. Palenzona, M.  Tropeano, M. R. Cimberle, C. Ferdeghini, R. Valle,
  M. Putti, A. S. Siri, Supercond. Sci. Technol. {\bf 21}, 095017
  (2008).

\bibitem{Lee08Struct} C.-H. Lee, A. Iyo, H. Eisaki, H. Kito,
  M. T. Fernandez-Diaz, T. Ito, K. Kihou, H. Matsuhata, M. Braden, and
  K. Yamada, J. Phys. Soc.  Jpn. {\bf 77}, 083704 (2008).

\end{thebibliography}
\end{document}